\documentclass[pra,amsmath,aps,floatfix,twocolumn,showpacs]{revtex4}

\usepackage{graphicx}
\usepackage{color}
\usepackage{marvosym}
\usepackage{wasysym}
\usepackage{pifont}
\usepackage{amssymb}
\usepackage{hyperref}

\renewcommand{\r}{{\bf r}}
\newcommand{\la}{\langle}
\newcommand{\ra}{\rangle}
\newcommand{\vect}[1]{\mathbf{#1}}

\newcommand{\mf}{m_{\scriptscriptstyle F}}

\begin{document}

\title{Stability of coreless vortices in ferromagnetic spinor
  Bose-Einstein condensates}
\author{V.~Pietil\"a,$^{1}$ M.~M\"ott\"onen,$^{1,2}$
    and S.~M.~M.~Virtanen$^{1}$ }
\affiliation{$^{1}$Laboratory of Physics, Helsinki University of
  Technology, P.O.~Box 4100,
FI-02015 TKK, Finland. \\
$^{2}$Low Temperature Laboratory, Helsinki University of Technology,
  P.O.~Box 3500,
FI-02015 TKK, Finland.}
\date{\today}

\begin{abstract}
We study the energetic and dynamic stability of coreless vortices in
nonrotated spin-1 Bose-Einstein condensates, trapped with a
three-dimensional optical potential and a Ioffe-Pritchard field. The stability
of stationary vortex states is investigated by solving the corresponding
Bogoliubov equations. We show that the quasiparticle excitations
corresponding to axisymmetric stationary states can be taken to be eigenstates
of angular momentum in the axial direction.
Our results show that coreless vortex states can occur
as local or global minima of the condensate energy or become energetically or
dynamically unstable depending on the parameters of the Ioffe-Pritchard field.
The experimentally most relevant coreless vortex state
containing a doubly quantized vortex in one of the hyperfine spin components
turned out to have very non-trivial stability regions, and especially a
quasiperiodic dynamic instability region which corresponds to splitting of
the doubly quantized vortex.

\end{abstract}

\hspace{5mm}


\maketitle

%%%%%%%%%%%%%%%%%%%%%%%%%%%%%%%%%%%%%%%%%%%%%%%%%%%%%%%%%%%%%%%%%%%%%%%%%%%%%%%
\section{Introduction}\label{intro}
%%%%%%%%%%%%%%%%%%%%%%%%%%%%%%%%%%%%%%%%%%%%%%%%%%%%%%%%%%%%%%%%%%%%%%%%%%%%%%%

Development of optical trapping techniques for alkali atoms has enabled
experimental studies of dilute atomic Bose-Einstein condensates (BECs) with
spin degrees of freedom \cite{stamper-kurn:1998,barrett:2001}. For these
systems, the order parameter is a spinor field which can exhibit a rich variety
of different topological textures ranging from coreless vortices
\cite{ho:1998,ohmi:1998} to monopoles
\cite{stoof:2001,martikainen:2002} and Skyrmion-type configurations
\cite{khawaja_nature:2001,khawaja_pra:2001,zhai:2003}. In scalar BECs, a
vortex is always fully characterized by the phase winding about the
vortex core, whereas in spinor BECs the characteristics of a vortex are
determined by winding numbers of different components as well as the core
polarization -- for coreless vortices the superfluid velocity is non-divergent
at the vortex core and hence the vortex core can be polarized.

Coreless vortices are topologically unstable, i.e., they can be
continuously deformed to a uniform texture. Thus their
existence as stable states typically requires
the presence of additional forces such as interactions at large
distances from the vortex core or external fields which impose nontrivial
asymptotic boundary conditions \cite{salomaa:1987,vollhardt:1990}.
Coreless vortices such as the Mermin-Ho vortices in spinor BECs are
analogous to those in superfluid $^3$He-A \cite{ho:1998}, in
which they appear as equilibrium objects if the system is rotated
externally. Thus it is natural to assume that such objects would be generated
also in rotated gaseous condensates. It has indeed been theoretically
confirmed that also in these systems the Mermin-Ho vortices are energetically
stable for certain values of the trap rotation frequency and
magnetization~\cite{mizushima_prl:2002}.

\begin{figure}[h]
\centering
\includegraphics[scale=0.2]{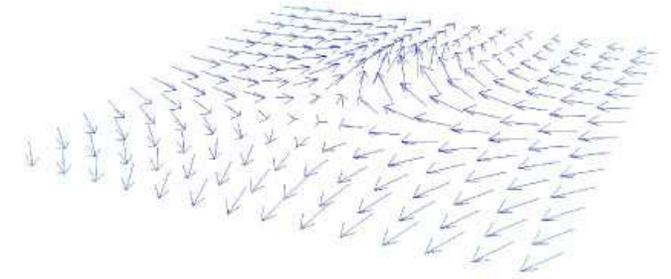}
\caption{\label{texture} (Color online) One possible spin texture of
  a coreless vortex in pancake-shaped condensate in the presence of the
  Ioffe-Pritchard field. Spin of the condensate tends to align with the
  external field.}
\end{figure}

Manipulating spinor condensates with external magnetic fields has been in
vogue among both theorists and experimentalist during the recent years
\cite{ho:1996,leanhardt:2002,leanhardt:2003,bulgakov:2003,zhang:2007}.
Topological phase engineering by time-dependent external magnetic fields has
been successfully used to create vortex structures
\cite{nakahara:2000,isoshima_1:2000,ogawa:2002,mottonen:2002,leanhardt:2002}.
Recently, Leanhardt {\it et al.} succeeded in creating a
coreless vortex in a $F=1$ spinor condensate in a Ioffe-Pritchard (IP)
magnetic trap~\cite{pritchard:1983} by adiabatically switching off the
magnetic bias field along the  trap axis \cite{leanhardt:2003}.
The ground state phase diagram corresponding to the IP~field combined with an
optical confinement potential has been computed  for a condensate uniform in
the direction of the vortex axis~\cite{bulgakov:2003}, and it shows that the
IP~field renders the ground state of the system to be a coreless vortex.
This is due to the tendency of the spin to align with the external field,
which leads to formation of a coreless vortex. An example of the spin texture
in an IP~field is shown in Fig.~\ref{texture} in the pancake-shaped geometry of
Ref.~\cite{bulgakov:2003}. Zhang {\it et al.}~\cite{zhang:2007} found that
within an adiabatic approximation~\cite{ho:1996}, the difference of atomic
spatial angular momentum and hyperfine spin is conserved in the IP~field. This
conservation law implies that the ground state of the condensate carries a
persistent current with definite winding numbers.

In the previous studies~\cite{bulgakov:2003,zhang:2007} only the global
energetic stability of coreless vortex states was considered. However, the
coreless vortices created in the experiment~\cite{leanhardt:2003} contain a
doubly quantized vortex in one of the hyperfine spin states, and it is of
interest to find out whether such coreless vortex states inherit the dynamic
instability of doubly quantized vortices in scalar condensates
\cite{pu:1999,mottonen:2003,huhtamaki_a:2006,munoz_mateo:2006,huhtamaki_b:2006,lundh:2006}.
Furthermore, the topological phase imprinting
methods~\cite{nakahara:2000,isoshima_1:2000,ogawa:2002,mottonen:2002,leanhardt:2002}
used to create doubly quantized vortices involve adiabatic inversion of the
bias field in the presence of the IP~field. The intermediate state in this
process is always a coreless vortex which for certain field parameters is the
ground state of the system but otherwise its stability properties are so far
unknown. Hence the study of stability of coreless vortices can also give
valuable insight to the process of vortex creation using topological phase
engineering.

In this paper, we analyze further the vortex stability phase diagram of
ferromagnetic spinor condensates in the presence of a IP~field
and an optical trap potential by computing the quasiparticle spectra
from the Bogoliubov equations in finite three-dimensional trap geometries, for
oblate, spherical, and prolate condensates. We show that different coreless
vortex states are locally energetically stable for a wide range of
IP~field configurations, although for other parameter values
dynamic and energetic instabilities may occur. Quasiparticle states
corresponding to these instabilities reveal that the dynamic instability of
coreless vortices containing doubly quantized vortices in one of the hyperfine
spin components is similar to the dynamic instability of the doubly quantized
vortices in scalar condensates, that is, the doubly quantized vortices tend to
split due to the instability.

%%%%%%%%%%%%%%%%%%%%%%%%%%%%%%%%%%%%%%%%%%%%%%%%%%%%%%%%%%%%%%%%%%%%%%%%%%%%%%
\section{Mean field theory}\label{mean}
%%%%%%%%%%%%%%%%%%%%%%%%%%%%%%%%%%%%%%%%%%%%%%%%%%%%%%%%%%%%%%%%%%%%%%%%%%%%%%

In the mean-field approximation, we describe the condensate formed
by weakly interacting ultra-cold bosonic atoms in the $z$-quantized
basis $|F=1,\mf\ra$, $\mf = -1,0,1$, with a spinor field $\psi =
(\psi_1\,\,\psi_0\,\,\psi_{-1})^T$ and the free energy functional of the form
\cite{ho:1998,ohmi:1998}
\begin{align}
\label{h_3comp}
\mathcal{E}[\psi]  = \int
 \mathrm{d}\r\bigg[&\frac{\hbar^2}{2m}|\nabla\psi(\r)|^2
 + U(\r)|\psi(\r)|^2 - \mu|\psi(\r)|^2 \nonumber \\
 &+\frac{c_0}{2}|\psi(\r)|^4 +\frac{c_2}{2} |\vect{S}(\r)|^2 \bigg],
\end{align}
where the spin density is given by
\begin{equation}
\label{spin_density}
\vect{S}(\r) = \sum_{a,b}\psi^{*}_a(\r)\vect{F}_{ab}\psi_b(\r),
\end{equation}
and the angular momentum matrices $\vect{F} = (F_x\,\,F_y\,\,F_z)^T$ are the
usual generators of the spin rotation group $SO(3)$ with matrix representations
%\begin{align*}
\begin{equation*}
F_x = \frac{1}{\sqrt{2}}\begin{pmatrix}  0 & 1 & 0 \\
                                         1 & 0 & 1 \\
                                         0 & 1 & 0
                              \end{pmatrix}, \quad
F_y = \frac{i}{\sqrt{2}}\begin{pmatrix}  0 & -1 & 0 \\
                                         1 & 0 & -1 \\
                                         0 & 1 & 0
                              \end{pmatrix}, \quad \\
F_z = \begin{pmatrix}  1 & 0 & 0 \\
                       0 & 0 & 0 \\
                       0 & 0 & -1
            \end{pmatrix}.
\end{equation*}
%\end{align*}
Above, $m$ is the mass of the atoms, $\mu$ the chemical potential, and
$c_0$, $c_2$ are the coupling constants related to $s$-wave scattering
lengths in different total hyperfine spin channels \cite{ho:1998}.
Depending on whether the interaction coupling constant $c_2$ is positive
or negative, the condensate is either antiferromagnetic or ferromagnetic,
respectively. Klausen {\it et al.} \cite{klausen:2001} have shown
that $^{87}$Rb is ferromagnetic, whereas $^{23}$Na atoms realize an
antiferromagnetic condensate~\cite{stenger:1998}.
We take the confining potential $U(\r)$ created by an external optical field
to be the axisymmetric harmonic potential
\begin{equation*}
U(\r) = \frac{1}{2}(\omega_r^2 r^2 + \omega_z^2 z^2) = \frac{\omega_r^2}{2}(
r^2 + \lambda^2 z^2),
\end{equation*}
where $\lambda = \omega_z/\omega_r$.

In the presence of an additional
external magnetic field $\vect{B}(\r)$, we take also into
account the linear Zeeman term
$\mu_{\scriptscriptstyle \mathrm{B}} g_{\scriptscriptstyle
  \mathrm{F}}\vect{B}(\r)\cdot\vect{S}(\r)$ in the energy functional. The
constant $\mu_{\scriptscriptstyle\mathrm{B}}$ is the Bohr magneton and
$g_{\scriptscriptstyle \mathrm{F}}$ is the Land\'{e} $g$-factor. The free
energy becomes in this case
\begin{equation}
\mathcal{F}[\psi] = \mathcal{E}[\psi] + \mu_{\scriptscriptstyle \mathrm{B}}
g_{\scriptscriptstyle
  \mathrm{F}}\int\mathrm{d}\r\,\vect{B}(\r)\cdot\vect{S}(\r),
\end{equation}
and the magnetic field is in our analysis of the
Ioffe-Pritchard form
\begin{equation*}
\vect{B}(\r) = B_{\perp}(x\hat{\vect{x}}
-y\hat{\vect{y}}) + B_z\hat{\vect{z}}.
\end{equation*}
Stationary states of the condensate satisfy $\delta \mathcal{F}[\psi]/\delta\psi = 0$,
which yields the Gross-Pitaevskii (GP) equation
\begin{equation}
\label{gp}
\mathcal{H}[\psi]\psi = \mu\psi,
\end{equation}
where the non-linear operator $\mathcal{H}[\psi]$ is given by
\begin{align*}
\mathcal{H}[\psi] =
&-\frac{\hbar^2}{2m}\nabla^2+U(\r) +\mu_{\scriptscriptstyle \mathrm{B}}
g_{\scriptscriptstyle
  \mathrm{F}}\vect{B}(\r)\cdot\vect{F}+c_0|\psi(\r)|^2  \nonumber \\
&+c_2 \vect{F}\cdot\vect{S}(\r).
\end{align*}

The quasiparticle spectrum corresponding to a given stationary state can be
solved from the generalized Bogoliubov equations
\cite{mizushima_prl:2002,isoshima:2000}
\begin{equation}
\label{bogo}
\mathcal{D}\begin{pmatrix} u_q(\r) \\ v_q(\r) \end{pmatrix}
= \hbar\omega_q \begin{pmatrix} u_q(\r) \\ v_q(\r) \end{pmatrix},
\end{equation}
where $u_q = (u_{q,1}^{}\,\,u_{q,0}^{}\,\,u_{q,-1}^{})^T$ and $v_q = (v_{q,1}^{}\,\,v_{q,0}^{}\,\,v_{q,-1}^{})^T$
are the quasiparticle amplitudes and the operator
$\mathcal{D}$ is defined as
\begin{equation*}
\mathcal{D} = \begin{pmatrix} A & -B \\
                              B^* & -A^*
              \end{pmatrix},
\end{equation*}
such that the components of the $A$ and $B$ operators are
\begin{align}
%\label{A}
A_{ij} = &\bigg(-\frac{\hbar^2}{2m}\nabla^2+U(\r) -\mu
\bigg)\delta_{ij}  +
\mu_{\scriptscriptstyle \mathrm{B}}
g_{\scriptscriptstyle
  \mathrm{F}}\sum_{\alpha}B_{\alpha}(\r)(F_{\alpha})_{ij} \nonumber \\
+&c_0\left\{\sum_{k} |\psi_k(\r)|^2\delta_{ij}+\psi_i(\r)\psi_j^*(\r)\right\} \nonumber \\
+&c_2\sum_{\alpha}\sum_{k,\,l}[(F_{\alpha})_{ij}(F_{\alpha})_{kl}+(F_{\alpha})_{il}(F_{\alpha})_{kj}]\psi_k^*(\r)\psi_l(\r), \nonumber
 \\
%\label{B}
B_{ij} = &
 c_0\psi_i(\r)\psi_j(\r)+c_2\sum_{\alpha}\sum_{k,\,l}(F_{\alpha})_{ik}(F_{\alpha})_{jl}\psi_k(\r)\psi_l(\r), \nonumber
\end{align}
for $i,j\in\{1,0,-1\}$.
Due to the conjugate symmetry of the Bogoliubov equations, we may concentrate
only on the quasiparticle modes for which the quadratic form
$\int \mathrm{d}\r\,(|u_q(\r)|^2- |v_q(\r)|^2)$ is non-negative.
The quasiparticle spectrum determines the stability properties
of the corresponding stationary state: If the quasiparticle spectrum contains
excitations with non-real frequencies $\omega_q$, the system is dynamically
unstable and even small initial perturbations can render the population of
these modes to grow exponentially in time even in the absence of
dissipation. On the other hand, if there exists modes with negative
eigenfrequencies, the state is energetically unstable, and in the presence of
dissipation the condensate can lower its energy by transferring particles
from the condensate state to such anomalous modes. Vice versa,
if all the quasiparticle eigenfrequencies are positive, the
state is locally energetically and dynamically stable, and should
be long-living and robust against small perturbations even in the
presence of dissipational mechanisms.

%%%%%%%%%%%%%%%%%%%%%%%%%%%%%%%%%%%%%%%%%%%%%%%%%%%%%%%%%%%%%%%%%%%%%%%%%%%%%%%
\section{Axisymmetric vortex states}\label{axial}
%%%%%%%%%%%%%%%%%%%%%%%%%%%%%%%%%%%%%%%%%%%%%%%%%%%%%%%%%%%%%%%%%%%%%%%%%%%%%%%

Within an adiabatic approximation, that is, assuming that hyperfine
spins of the atoms remain aligned (or antialigned) with the local magnetic
field, the ground state of the condensate  in the IP~field  is
axisymmetric~\cite{zhang:2007}. However, nonadiabatic effects
are important to some extent, since in Ref.~\cite{bulgakov:2003} the ground
state phase diagram was shown to contain  a non-axisymmetric vortex
state in the antiferromagnetic case. On the other hand, we have
verified with fully three-dimensional (3D) computations, i.e.~without
setting any symmetry restrictions to the condensate wavefunction, that
for ferromagnetic condensates with nonzero radial IP~field strength,
the ground state is always axisymmetric for all the parameter values
considered. Consequently, we restrict to consider only configurations
that are axisymmetric.

In cylindrical coordinates $\r = (r,\varphi,z)$, axially symmetric vortex
states are of the form
\begin{equation}
\label{axisym_states}
\psi(\r)=
\begin{pmatrix}
\psi_1(r,z)e^{i\kappa_1\varphi} & \psi_0(r,z)e^{i\kappa_0\varphi} &
\psi_{-1}(r,z)e^{i\kappa_{-1}\varphi}
\end{pmatrix}^T,
\end{equation}
where $\kappa_i\in\mathbb{Z}$
are the winding numbers of the three components---we
refer to states of this form as $\la \kappa_1, \kappa_0, \kappa_{-1}\ra$.
In addition, all physically measurable densities corresponding to
axisymmetric states have to be axisymmetric. When applied to
spin-density, this requirement combined with Eqs.~\eqref{spin_density}
and \eqref{axisym_states} implies the relation
\begin{equation}
\label{winding_numbers}
2\kappa_0 = \kappa_1 +\kappa_{-1},
\end{equation}
between the winding numbers. The angular dependence
of the IP~field sets additional restrictions for the
ground state to be axisymmetric: For the Zeeman energy term
$\mu_{\scriptscriptstyle \mathrm{B}} g_{\scriptscriptstyle
  \mathrm{F}}\vect{B}(\r)\cdot\vect{S}(\r)$
to be rotationally symmetric the winding numbers of the condensate
state have to satisfy the additional constraints
\begin{equation}
\label{linear_zeeman}
\kappa_0=\kappa_1-1=\kappa_{-1}+1.
\end{equation}
One notes that these latter relations imply also the relation
in Eq.~\eqref{winding_numbers}, and are thus more restrictive. We have
verified with  fully 3D computations without any symmetry assumptions that the
ground state of system indeed satisfies the restrictions given in the
Eqs.~\eqref{axisym_states}--\eqref{linear_zeeman}. The states with the lowest
angular momenta satisfying Eqs.~\eqref{winding_numbers} and
\eqref{linear_zeeman} are $\la 2,1,0\ra$, $\la 1,0,-1\ra$, and $\la
0,-1,-2\ra$ which are coreless vortex states. The numerical results indeed
show that the ground state in the presence of a strong enough IP~field
contains a coreless vortex.

The coreless vortices in the states $\la 2,1,0\ra$ and $\la
0,-1,-2\ra$ are ferromagnetic in the sense that the spin of the condensate is
aligned (or antialigned) with the external field also in the core
region. Furthermore, $\la 2,1,0\ra$ and $\la 0,-1,-2\ra$ are equivalent since
they differ only by inversion of the spin quantization axis. On the other
hand, $\la 1,0,-1\ra$ is polar in the sense that at the vortex core
$\vect{S}(\r)$ vanishes~\cite{ho:1998,isoshima:2000}. The difference in the
spin texture turns out to be significant for the phase diagram and the local
energetic stability of the vortex state. Due to vanishing spin density at the
vortex core, the spin texture of the polar vortex $\la 1,0,-1\ra$ can align
with the IP~field equally well for both positive and negative values of the
bias field $B_z$.

In investigating the stability properties of stationary states
by solving the Bogoliubov equations, it is to be noted that apart from
the possible degeneracy of spectrum, axisymmetric
states can have non-axisymmetric quasiparticle excitations also due to the
fact that $\mathcal{D}$ does not in general commute with the angular momentum
operator $\hat{L}_z = -i\hbar\partial_{\varphi}$. We can,
however, show that if the winding numbers of different components satisfy
Eq.~\eqref{linear_zeeman}, then there exists a unitary transformation
$\mathcal{U}$ such that the transformed Bogoliubov operator
$\widetilde{\mathcal{D}} = \mathcal{U}^{\dagger}_{}\mathcal{D}\mathcal{U}$
commutes with $\hat{L}_z$. The unitary transformation $\mathcal{U}$ is in this
case given by
\begin{equation*}
\mathcal{U} = \begin{pmatrix}
                \mathcal{U}_0 & 0 \\
                0 & \mathcal{U}^{\dagger}_0
                \end{pmatrix},
\,\,\,\,\mathrm{where}\,\,\,\, \mathcal{U}_0^{} = \begin{pmatrix}
                e^{i\kappa_1\varphi} & 0 & 0 \\
                0 & e^{i\kappa_0\varphi} & 0 \\
                0 & 0 & e^{i\kappa_{-1}\varphi}
                \end{pmatrix},
\end{equation*}
for the Bogoliubov operator  $\mathcal{D} = \mathcal{D}[\psi]$ corresponding
to an axisymmetric state $\la \kappa_1,\kappa_0,\kappa_{-1}\ra$. Now
$[\widetilde{\mathcal{D}},\hat{L}_z] = 0$ follows from the
condition stated in Eq.~\eqref{linear_zeeman}. Hence the eigenstates of
$\widetilde{\mathcal{D}}$ can be chosen to be eigenstates of
$\hat{L}_z$. Writing the eigenvalue equation $\widetilde{\mathcal{D}}w = \eta
w$ in the form
$\mathcal{U}\widetilde{\mathcal{D}}\mathcal{U}^{\dagger}_{}\mathcal{U}w =
\eta\mathcal{U}w$ and taking into account that
$\mathcal{U}\widetilde{\mathcal{D}}\mathcal{U}^{\dagger}_{} = \mathcal{D}$, we
observe that the quasiparticle amplitudes in the original
Bogoliubov equation~\eqref{bogo} can be taken to be of the form
\begin{align}
u_{q,j}^{}(\r) & = u_{q,j}^{}(r,z)e^{i(\kappa_{q}^{}+\kappa_j^{})\varphi},\qquad \\
v_{q,j}^{}(\r) & = v_{q,j}^{}(r,z)e^{i(\kappa_{q}^{}-\kappa_j^{})\varphi},\qquad j = -1,0,1,
\end{align}
where $\kappa_{q}$ is an angular momentum quantum number of the
excitation. In addition to this analytical argument, we have verified
numerically without any symmetry assumptions that for the pancake-shaped
condensates with $\lambda=\omega_z/\omega_r\gg1$, all the low-energy
quasiparticle states are axisymmetric.

%%%%%%%%%%%%%%%%%%%%%%%%%%%%%%%%%%%%%%%%%%%%%%%%%%%%%%%%%%%%%%%%%%%%%%%%%%%%%%%
\section{Numerical results}\label{num_res}
%%%%%%%%%%%%%%%%%%%%%%%%%%%%%%%%%%%%%%%%%%%%%%%%%%%%%%%%%%%%%%%%%%%%%%%%%%%%%%%

In the following we consider ferromagnetic condensates which can be
realized, e.g., with $^{87}$Rb atoms. For the scattering lengths of
$^{87}$Rb, the estimate of van Kempen {\it et al.}~\cite{vanKempen:2002}
implies the ratio  $c_2/c_0 \sim -0.005$. In this study we use the value
$c_2/c_0 = -0.02$ which was also used in the previous theoretical
studies of Mizushima {\it et
  al.}~\cite{mizushima_prl:2002,mizushima_pra:2002}.
In the numerical calculation we use dimensionless units in which $\tilde{c}_0
= c_0 mN/a_r\hbar$, where the characteristic length scale of the trap is
given by  $a_r = \sqrt{\hbar/m\omega_r}$. We adopt $\tilde{c}_0 =
10000$, which corresponds, for example, a condensate of $N = 8.5\times
10^5$ $^{87}$Rb atoms trapped such that $\omega_r = 2\pi\times
200$ Hz. We take the Land\'{e} $g$-factor to be $g_{\scriptscriptstyle
  \mathrm{F}} = -\frac{1}{2}$ which indicates that the hyperfine spine tends
to align with an external field. We have searched for solutions of the
GP~equation and the Bogoliubov equations by using finite-difference
discretization combined with relaxation methods and the implicitly
restarted Arnoldi method implemented in the ARPACK numerical
library~\cite{arpack}. In the numerical calculations, we also use the
conjugate symmetry of the Bogoliubov equations and consider only
positive $\kappa_q$, which slightly reduces the numerical effort.

\begin{figure}[h]
%\centering
\includegraphics[scale=0.75]{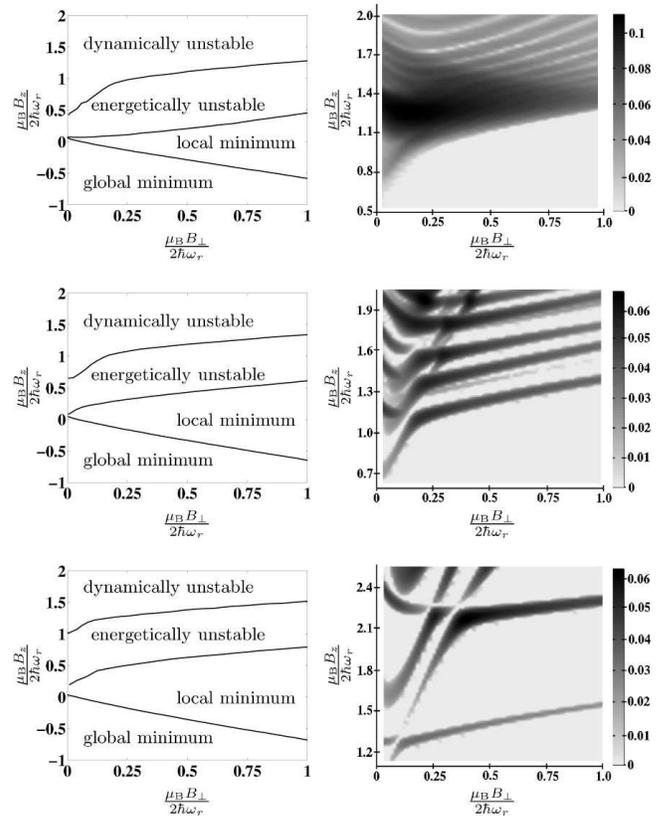}
\caption{\label{phase_diagrams} Stability phase diagram for the axially
  symmetric $\la 2,1,0\ra$ vortex state. On the top $\lambda =
  0.2$, in the middle $\lambda = 1.0$, and at the bottom $\lambda = 5.0$. The
  left panels show the different stability phases and the right panels the
  maximum value of $|\mathrm{Im}\{\omega_q/\omega_r\}|$ for each point in the
  $(B_{\perp},B_z)$ plane as a grayscale plot: bright regions correspond to
  dynamically stable regions, and dark ones dynamically unstable. In the
  region denoted as ``local minimum'' $\la 2,1,0\ra$  is
  a local minimum of the mean field energy and the polar state $\la 1,0,-1\ra$
  is the ground state of the system. The dynamic instability region
  indicates the section of the parameter space where dynamic instability
  modes can appear.}
\end{figure}

\begin{figure}[h!]
\centering
\includegraphics[scale=0.975]{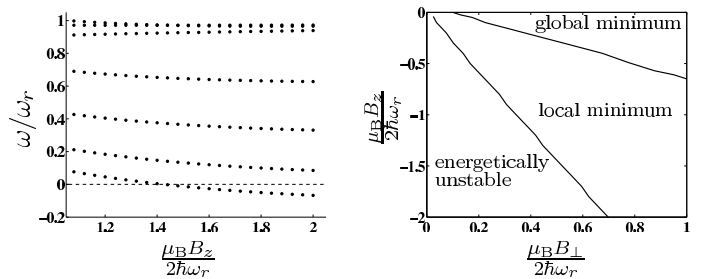}
\caption{\label{polar} Stability phase diagram of the polar
  vortex $\la 1, 0,-1\ra$  (right panel) and the lowest energy excitations
  as a function of $B_z$ for fixed
  $\mu_{\scriptscriptstyle\mathrm{B}}B_{\perp}/2\hbar\omega_r = 0.5$ (left
  panel). The phase diagram is symmetric with respect to reversing the
  sign of $B_z$. }
\end{figure}

Numerical computations showed that the ground state configuration
for non-vanishing perpendicular IP~field $B_{\perp}$
is always one of the coreless vortex states $\la 2,1,0\ra$,
$\la 1,0,-1\ra$ or $\la 0, -1, -2\ra$. Figure~\ref{phase_diagrams}
shows the computed stability phase diagrams displaying the dynamic and
energetic stability/instability regions for the state $\la 2,1,0\ra$ as a
function of the IP~field parameters. The corresponding phase diagram for the
state $\la 1,0,-1\ra$ is shown in Fig.~\ref{polar}. The results have been
computed for the trap asymmetry parameter values $\lambda= 0.2$, $1.0$, and
$5.0$, corresponding to prolate, spherical, and oblate geometries,
respectively. The phase diagram for the state $\la 0, -1, -2\ra$ is the same
as for $\la 2,1,0\ra$ if the sign of the bias field $B_z$ is reversed.

\begin{figure*}[t!]
\centering
\includegraphics[scale=0.88]{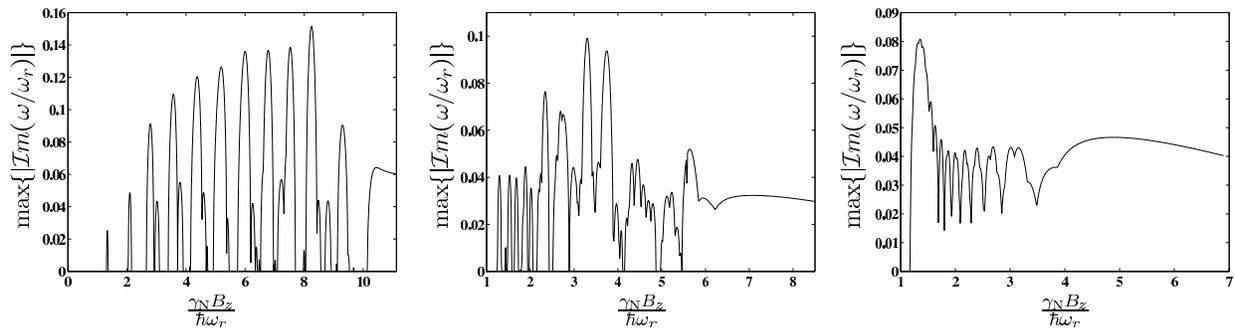}
\caption{\label{fixed_b} Quasiperiodic structure of the dynamic instability as
  a function of $B_z$  for the axially symmetric
  $\la 2,1,0\ra$ vortex and $\frac{\mu_{\scriptscriptstyle\mathrm{B}}B_{\perp}}{2\hbar\omega_r} = 0.7$. Left panel: $\lambda =
  5.0$;  middle: $\lambda = 1.0$; and right panel: $\lambda = 0.2$.}
\end{figure*}

From the diagrams one infers that the ferromagnetic state $\la 2,1,0\ra$ and
the polar state $\la 1,0,-1\ra$ can co-exist as local energy minima for
suitable IP field parameters. The transition between the ferromagnetic and
polar ground states is determined by the relative magnitude of the IP field
components: for $B_{\perp} \gg |B_z|$ the ferromagnetic core of the vortex
states $\la 2,1,0\ra$ and $\la 0,-1,-2\ra$ becomes   energetically
unfavorable, rendering the polar state $\la 1,0,-1\ra$ the global minimum of
energy, and vice versa for $|B_z| \gg B_{\perp}$.

The negative energy anomalous modes indicating energetic instability occur
for quantum numbers $\kappa_q = \pm 1$ for the polar state $\la
1,0,-1\ra$ and for $\kappa_q = \pm 1, \pm 2$ for the
ferromagnetic state $\la 2,1,0\ra$.
The $\la 1,0,-1\ra$ state turned out to be dynamically stable for
all the parameter values investigated, but the states
$\la 2,1,0\ra$ and $\la 0, -1, -2\ra$ have complicated dynamic
instability regions. The dynamic instability modes occur only for $\kappa_q =
\pm 2$. Figure~\ref{phase_diagrams} shows the greyscale plots of
the maximum imaginary parts of the Bogoliubov eigenfrequencies for
the state $\la 2,1,0\ra$
as functions of the IP~field parameters for three trap geometries.
We note that the regions marked as dynamically unstable
in Fig.~\ref{phase_diagrams} contain narrow stripe-like patterns in which the
vortex state is in fact dynamically stable. However, especially
for the prolate geometry these stripes are very narrow and probably
experimentally indistinguishable. The stripe-like quasiperiodic structure
of the dynamic instability can be observed more clearly in
Fig.~\ref{fixed_b}, in which the maximum imaginary part is
plotted  for a fixed value of $B_{\perp}$. One observes that the magnitude of
the largest imaginary part oscillates markedly
before it saturates for strong enough bias fields $B_{z}$.

\begin{figure}[h!]
\centering
\includegraphics[scale=0.9]{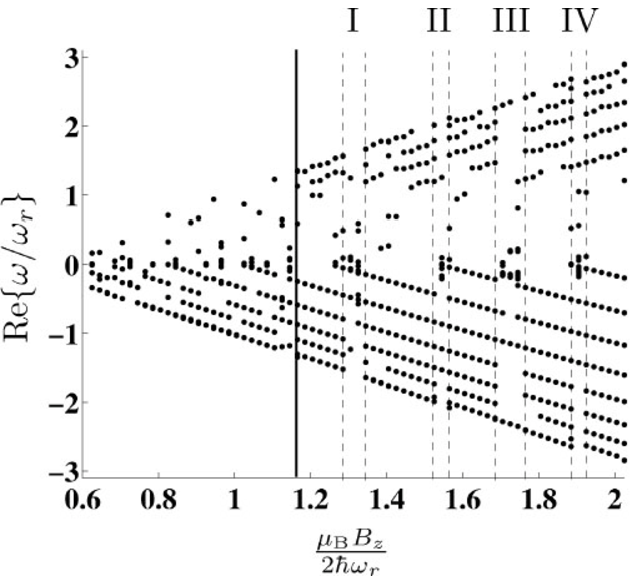}
\caption{\label{ferro_spectrum} The negative energy excitations and the
  corresponding positive energy excitations with $\kappa_q = \pm 2$ for the
   ferromagnetic vortex state $\la 2,1,0\ra$. Here
   $\mu_{\scriptscriptstyle\mathrm{B}
  }B_{\perp}/2\hbar\omega_r = 0.5$ and $\lambda = 1.0$ are fixed. The dynamic
   instabilities start to occur for
   $\mu_{\scriptscriptstyle\mathrm{B}}B_z/2\hbar\omega_r \gtrsim 1.16$ which
   is denoted by the solid line. The intervals denoted by the dashed lines and
   the Roman numerals I, II, III, and IV correspond to regions of
   values of $B_z$ for which the $\la 2,1,0\ra$ vortex state is dynamically stable
   (cf.~Fig.~\ref{phase_diagrams}).}
\end{figure}

Figure~\ref{phase_diagrams} shows that the $\la 2,1,0\ra$ vortex tends to
become dynamically more stable with increasing $\lambda$, i.e.,
the vortex is generally more stable in the pancake shaped geometry than in
the cigar shaped one. This is due to the suppression of the density of
low-energy excitations in the limit of tight confinement in the
$z$ direction. However, the dynamic instability persists even for
$\lambda=\omega_z/\omega_r\gg 1$. The energetic stability of the coreless
vortices is intimately related to dynamic instability since it has been
shown in Ref.~\cite{lundh:2006} that dynamic instabilities in scalar
condensates are formed when a negative and a positive energy excitation are
in resonance. Similar phenomenon takes place also in spinor condensates,
and can be observed in this case by inspecting the spectrum of the
quasiparticle energies for $\kappa_q = \pm 2$ shown in
Fig.~\ref{ferro_spectrum}. For $\lambda = 1.0$ and
$\mu_{\scriptscriptstyle\mathrm{B}}B_{\perp}/2\hbar\omega_r = 0.5$,
the dynamic instability may occur for parameter values
$\mu_{\scriptscriptstyle\mathrm{B}}B_z/2\hbar\omega_r \gtrsim 1.16$. From
Fig.~\ref{ferro_spectrum} one observes that some positive energy excitations
are missing in the region
$\mu_{\scriptscriptstyle\mathrm{B}}B_z/2\hbar\omega_r \gtrsim 1.16$ due to the
resonance. We note that the regions in which the $\la 2,1,0\ra$ vortex is
dynamically stable seem to be
manifested by the absence of certain positive and negative energy excitations.
For the polar vortex $\la 1,0,-1\ra$ the excitation
spectrum is much simpler containing only one anomalous mode. The stability
phase diagram for the polar state $\la 1,0,-1\ra$ in spherical geometry is
shown in Fig.~\ref{polar} together with the lowest quasiparticle energies.

For the ferromagnetic vortex state $\la 2,1,0\ra$, the unit vector
field $\hat{\vect{n}}(\r) = \vect{S}(\r)/|\vect{S}(\r)|$ is well-defined
everywhere inside the cloud. Thus the spin texture of this state can be mapped
to a simply connected subset $\mathcal{S}\subset S^2$ and
the perpendicular part of the IP~field prevents the subset $\mathcal{S}$ from
shrinking to a point. The surface $\mathcal{S}'$ corresponding to
the spin texture of the polar vortex $\la 1,0,-1\ra$ contains a hole due to
vanishing $\vect{S}(\r)$ at the vortex core. Thus the two spin textures are
topologically inequivalent and there is a topological phase transition of the
ground state between the two textures occurring at the boundary between the
regions of global and local stability in Fig.~\ref{phase_diagrams}.

\begin{figure}[h!]
\centering
\includegraphics[scale=0.90]{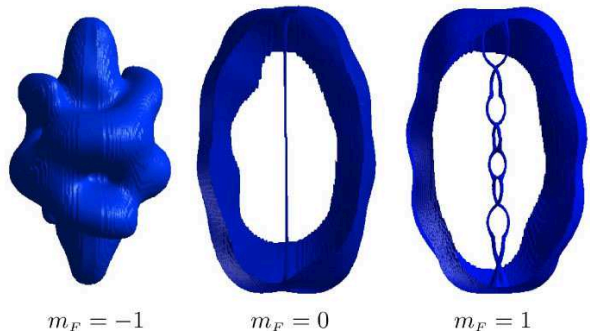}
\caption{\label{012_isosurf}  (Color online) Isosurfaces of particle
  densities in different spin components corresponding to vortex state
  $\la 2,1,0\ra$ and a slight excitation of a dynamic instability mode. The
  trap asymmetry parameter is $\lambda = 0.2$. Population of
  the dynamic instability mode is 1\% of the total number of particles. The
  isosurfaces correspond to values $|\Psi_1(\vect{r})|^2 = 2.6\cdot
  10^{-5}\,N/a_r^3$, $|\Psi_0(\vect{r})|^2 = 1.0\cdot 10^{-5}\,N/a_r^3$,
  and $|\Psi_{-1}(\vect{r})|^2 = 1.0\cdot 10^{-5}\,N/a_r^3$.}
\end{figure}

To investigate qualitatively the nature of the dynamic instability,
we consider states of the form
\begin{equation}
\label{perturbed_state}
\tilde{\psi}(\r) = \psi(\r) + \eta[u_q(\r) + v_q^*(\r)],
\end{equation}
where a slightly populated quasiparticle state
corresponding to the dynamic instability with the largest
imaginary frequency has been added to the corresponding stationary state.
Isosurfaces of particle densities in different hyperfine spin components of
such slightly perturbed $\la 2,1,0\ra$
vortex states are shown in Fig.~\ref{012_isosurf}. In the component
containing the doubly quantized vortex, one observes a helical vortex chain
structure, which is very similar to the one discovered in the numerical
calculations of Huhtam\"aki {\it  et al.}~\cite{huhtamaki_a:2006} for the
splitting times of doubly quantized vortices in scalar condensates.
Thus we observe that for the $\la 2,1,0\ra$ vortex, the dynamic instability
is essentially due to the splitting of the doubly quantized vortex in the $\mf
= +1$ component. Due to the helical structure of the splitted vortex
in this multicomponent case, the splitting of the doubly quantized
vortex is accompanied with a spiral flow about the $z$-axis of the
vortex free component $\mf = -1$.

Due to the Zeeman energy term, the average spin $\vect{S}(\r)$ tends to align
with the local magnetic field. However, in the vicinity of the vortex
core the average spin may deviate from the direction of
the local magnetic field without costing too much energy. The spin texture of
the state $\la 2,1,0\ra$
outside the vortex core points always along  the local magnetic field, but for
$B_z > 0$, $\vect{S}(\r)$ has opposite direction to the magnetic field. Since
the perpendicular part of the IP~field is topologically nontrivial with winding
number $-1$~\cite{leanhardt:2003,mermin:1979}, a continuous transformation of
the spin texture to energetically more favorable state cannot
be accomplished without spin rotations against the local magnetic
field. Thus the IP~field creates an energy barrier which
prevents the spin texture from unwinding itself via spin rotations.

%%%%%%%%%%%%%%%%%%%%%%%%%%%%%%%%%%%%%%%%%%%%%%%%%%%%%%%%%%%%%%%%%%%%%%%%%%%%%%
\section{Conclusions}\label{sec6}
%%%%%%%%%%%%%%%%%%%%%%%%%%%%%%%%%%%%%%%%%%%%%%%%%%%%%%%%%%%%%%%%%%%%%%%%%%%%%%

In conclusion, we have studied the local stability of coreless vortex states
in optically trapped ferromagnetic spinor $F=1$ BECs in the presence
of the Ioffe-Pritchard field. The ground states of the system turned
out to be axisymmetric and it was shown analytically that the
corresponding  quasiparticle states are also axisymmetric. By
computing numerically the quasiparticle spectra we have shown that
there can co-exist different coreless vortex states as local and
global minima of energy for a wide variety of different external field
configurations. The experimentally most tractable vortex
configuration, the coreless vortex state $\la 2,1,0\ra$, was shown to
possess a rich phase diagram in which the vortex transforms gradually
from a global minimum of energy to a dynamically unstable stationary
state as the bias field of the Ioffe-Pritchard trap is ramped up from
negative to positive bias. Based on these results one should be able
to realize experimentally robust coreless vortex states by loading the
atoms into an optical trap with an IP~field and allowing the
condensate to relax into the ground state. This method should also
enable experimental creation of the polar vortex state $\la
1,0,-1\ra$. Interesting questions for the future research would be to
investigate the phase diagram of the ground state and the local stability
of vortex states in an external Friedburg-Paul (hexapole) magnetic
field~\cite{hinds:2000,pu:2001}, and the exact decay mechanisms of energetically
unfavorable spin textures.

\begin{acknowledgments}

CSC-Scientific Computing Ltd (Espoo, Finland) is acknowledged for
computational resources and Academy of Finland for extensive financial
support. M.M.~thanks Finnish Cultural Foundation, V\"ais\"al\"a Foundation,
and Magnus Ehnrooth Foundation for financial support. V.P.~thanks the Jenny
and Antti Wihuri Foundation for financial support. J.A.M.~Huhtam\"aki is
appreciated for stimulating discussions.

\end{acknowledgments}

\bibliography{manuscript}

\end{document}